# Quantification of Collateral Supply with Local-AIF Dynamic Susceptibility Contrast MRI Predicts Infarct Growth

Mira M. Liu[1,2,*], Niloufar Saadat[3], Steven P. Roth[4], Marek A. Niekrasz[5], Mihai Giurcanu[6], Timothy J. Carroll[7], Gregory A. Christoforidis[8,9]

[1]Department of Radiology Medical Physics, University of Chicago, Chicago, IL, USA

[2]Current affiliation: Biomedical Engineering and Imaging Institute, Icahn School of Medicine at Mount Sinai, New York, NY, USA

[3]Department of Interventional Radiology, University of Chicago, Chicago, IL, USA

[4]Department of Anesthesiology, University of Illinois, Chicago, IL, USA

[5]Department of Surgery and Large Animal Studies, University of Chicago, Chicago, IL, USA

[6]Department of Statistics, University of Chicago, Chicago, IL, USA

[7]Department of Radiology Medical Physics, University of Chicago, Chicago, IL, USA

[8]Department of Interventional Radiology, University of Chicago, Chicago, IL, USA

[9]Current affiliation: Mount Carmel Health Systems, Columbus, OH, USA

**\*Corresponding Author:**
Mira M. Liu
Icahn School of Medicine at Mount Sinai
BioMedical Engineering and Imaging Institute
1470 Madison Ave, 1st Floor
New York, NY 10029
Tel:     (212) 824-8480
E-mail: mirabai.liu@mountsinai.org

**Grant Support**

The authors would like to acknowledge the support of the US National Institutes of Health R01-NS093908 (Carroll/Christoforidis) and The National Science Foundation DGE-1746045 (Liu).

**ABSTRACT**

**Background and purpose:** In ischemic stroke, leptomeningeal collaterals can provide compensatory blood flow to tissue at risk despite an occlusion, and impact treatment response and infarct growth. The purpose of this work is to test the hypothesis that local perfusion with an appropriate Local Arterial Input Function (AIF) is needed to quantify the degree of collateral blood supply in tissue distal to an occlusion.

**Materials and methods:** Seven experiments were conducted in a pre-clinical middle cerebral artery occlusion model. Magnetic resonance dynamic susceptibility contrast (DSC) was imaged and post-processed as cerebral blood flow maps with both a traditionally chosen single arterial input function (AIF) applied globally to the whole brain (i.e. "Global-AIF") and a novel automatic delay and dispersion corrected AIF (i.e. "Local AIF") that is sensitive to retrograde flow. Pial collateral recruitment was assessed from x-ray angiograms and infarct growth via serially acquired diffusion weighted MRI scans both blinded to DSC.

**Results:** The degree of collateralization at x-ray correlated strongly with quantitative perfusion determined using the Local AIF in the ischemic penumbra ($R^2=0.81$) compared to a traditionally chosen Global-AIF ($R^2=0.05$). Quantitative perfusion calculated using a Local-AIF was negatively correlated (less infarct progression as local perfusion increased) with infarct growth ($R^2 = 0.79$) compared to Global-AIF ($R^2=0.02$).

**Conclusions:** Local DSC perfusion with a Local-AIF is more accurate for assessing tissue status and degree of leptomeningeal collateralization than traditionally chosen AIFs. These findings support use of a Local-AIF in determining quantitative tissue perfusion with collateral supply in occlusive disease.

**Abbreviations:** MRI, magnetic resonance imaging; DSC, dynamic susceptibility contrast; PCS, pial collateral score; MCAO, middle cerebral artery occlusion; AIF, arterial input function; CBF, cerebral blood flow;

**Disclosure of potential conflicts of interest:**
The authors declare no conflicts of interest related to the content of this article.

## INTRODUCTION

In acute ischemic stroke where a major blood vessel feeding the brain becomes occluded, it is well known that development of compensatory blood flow through the leptomeningeal collateral arterial network (i.e., pial collateral supply) can provide significant blood supply and slow the growth of an infarction of tissue. Specifically, the leptomeningeal vessels can maintain blood supply to tissue at risk for an extended period despite an occlusion. While "door-to-needle time" currently drives ischemic stroke/triage/treatment protocols regardless of an individuals' collateral supply, neuroprotection/bridge therapies which enhance collateral supply for "drip-and-ship" cases, and "wake up strokes"[1] with unknown time of onset may benefit from knowledge of accurate pial collateral supply[2]. Further, the patient-to-patient variability of leptomeningeal collateral blood supply contributes to unwanted variability of perfusion-diffusion mismatch, a measure used to predict positive outcomes in patients undergoing thrombectomy[3]. As previous studies have observed a slowing of infarct growth with flow augmentation therapies[4,5], and that the success of flow augmentation may depend on collateralization[6,7], accurate quantification of robust collateral supply is needed to identify viable tissue for bridge therapies and provide more accurate prediction of infarct growth.

Leptomeningeal collateral blood supply can be estimated by x-ray digital subtraction angiography via late parenchymal enhancement, arterial input function (AIF)-independent single-photon emission computed tomography perfusion, computed tomography perfusion, and magnetic resonance perfusion[1] via prolonged Tmax[8] from dynamic susceptibility contrast (DSC). However, difficulties measuring absolute cerebral blood flow (CBF) with traditional DSC include highly variable compensatory vasodilation of cerebral blood volume (CBV)[9] and deconvolution of a single Global AIF overcalling perfusion deficits[10] suggesting it to be an inaccurate measure of blood supplied locally through the leptomeningeal collaterals.

This study investigates if DSC with an automatic local arterial input function (Local-AIF) may reflect compensatory blood supply to tissue at risk of infarction more accurately than traditional DSC deconvolution analysis (Global-AIF). Using a standard MRI acquisition sequence in a preclinical animal model of ischemic stroke[5,11-13] we post-processed DSC with both a traditional Global-AIF[14], and an automatically generated Local-AIF adjusted for delay and dispersion of the contrast bolus. We hypothesized that quantification of local perfusion with an appropriate Local Arterial Input Function (AIF) would correlate with degree of leptomeningeal collateral supply to a perfusion bed distal to an occlusion and better predict infarct growth.

**MATERIALS AND METHODS**

**Experimental Protocol:** All experiments were conducted using a preclinical canine model of ischemic stroke.[9,11,12] The two-day experimental protocol was approved by the [BLINDED FOR REVIEW] Institutional Animal Care and Use Committee and reported in compliance with ARRIVE guidelines. In this study, a series of 7 purpose bred adult canines (mean age = 4.1±2.9y, mean weight = 25.8±3.7kg, 6 female, 1 male) underwent permanent endovascular MCAO via embolic occlusion coils at the M1 segment. MCAO was then verified via selective ipsilateral and contralateral internal carotid and vertebral arteriography using a previously reported technique[5-7,12,13]. Vertebral and bilateral internal carotid arteriograms confirmed MCAO without involvement of other vessels and assessed pial collateral recruitment. Subjects underwent arteriography and all CBF assessment under general anesthesia.

Isoflurane (1% End-Tidal, 0.75 minimum alveolar concentration for canines), propofol infusion (6 mg/kg IV followed by continuous 0.1-0.2 mg/min), and intravenous rocuronium (0.4-0.6 mg/kg every 10-30 minutes) were used to maintain anesthesia with minimal influence on cerebral perfusion. Physiology was monitored via invasive blood pressure, End-Tidal $CO_2$, $O_2$ saturation, rectal temperature, heart rate, cardiac rhythm, arterial blood gases, glucose, electrolytes, and hematocrit. Pial collateral recruitment was quantified 30 minutes post-MCAO by assessing x-ray arteriographic images (OEC9800; General Electric Healthcare, Chicago, IL). Pial collateral recruitment was quantified from x-ray angiograms as a pial collateral score (PCS)[13] blinded to CBF. In short, the PCS assigns an ordinal score, from 1-11, based on the delay and extent of retrograde filling of arterial branches distal to the occluded artery, with 1 being minimal/no collateralization and 11 being retrograde reconstitution of the collateral network up to the occlusion. To dichotomize PCS, subjects with PCS ≤ 8 were considered "poor" recruitment, and PCS ≥ 9 considered "good" collateral recruitment, shown previously to indicate speed of infarction[13]. After deployment of the coil subjects were transported to the MRI scanner for advanced physiologic imaging. Battery operated pumps and ventilators were affixed to the specially designed transport cart to maintain physiologic parameters within acceptable ranges throughout transport. In addition, real-time physiology monitoring (Heart Rate, BP, EtCO2, PaCO2) were also displayed to the veterinary staff for the duration of the imaging to identify any changes that would bias results.

The animals then underwent MRI DSC perfusion and serial quantification of infarct volume by DTI and were euthanized either the same evening or the following day. Those with unanticipated events such as abnormal baseline MRI, intraprocedural intracranial vessel perforation, or excessive deviation in physiologic parameters during MCAO (presumably from an unanticipated procedural related event) were excluded from this study.

**MR Acquisition:** All MRI scans were performed on a 3T MRI scanner (Ingenia Philips, Cambridge, MA) with canines in a head-first, prone position using a 15 channel receive-only coil. DSC perfusion images were acquired 2.5 hours after occlusion (2D Gradient echo, T2*-weighted EPI, FOV=160 x160mm/matrix =176x176, 5 slices/6mm thick, TR/TE=500/30, 120 phases, total scan time = 60s) were post-processed to create parametric images of CBF. Rapid, 15s T1 maps were acquired using a 2D Inversion recovery Look Locker scan with a single-shot EPI readout (slice thickness = 6mm, FOV = 160 x 160 mm/matrix = 176 x 176)[9] for the T1-bookend method. Gadolinium (Gd)-based contrast agent (Multihance, Bracco, Princeton, NJ, USA) was injected in the forepaw followed by a saline flush (Gd: 3mL at 2mL/s, saline: 20mL at 2mL/s).

Diffusion tensor images (DTI) were acquired at 30-minute intervals to calculate mean diffusivity (MD) post-MCAO for measuring infarct volume growth over time. A stack of 50 2D DTI was prescribed to cover the entire head (slice thickness= 2mm FOV =128x128 mm/matrix = 128x128, TR/TE = 2993/83ms, FA = 90°, b-values = 0, 800 s/mm$^2$, 32 directions). MD, rather than ADC, maps were used as DTI was acquired to study separate aspects of this model, the analysis being beyond the scope of the current study.

**Traditional DSC Post-Processing:** DSC perfusion weighted images were post-processed with two methods; (1) using a traditional single AIF as input to deconvolution analysis globally, throughout the whole brain (Global-AIF) and (2) with a voxel-by-voxel, delay-and-dispersion corrected Local-AIF. In both cases, operators were blinded to the PCS. Images were calibrated to quantitative CBF (in ml/100g/min) and CBV (in ml/100g) using the quantitative T1-bookend method[15] where T1(in ms) changes in the parenchyma and blood-pool are input to a two-compartment model[16-18] that yields qCBF (in ml/100g/min). This approach to quantification has undergone extensive validation by direct comparison to reference standards (H2[O15] PET) and neutron capture microsphere deposition in the same animal model[9,10,19]. Bookend perfusion pulse sequence and post processing software was fully automated to minimize user bias[17,20] and is available for use[21,22].

For traditional DSC quantitative CBF (qCBF) the single Global AIF was chosen automatically based on a simultaneous assessment of early arrival time, narrow bolus, and large area under the concentration curve[23]. Relative CBV was calibrated to quantitative CBV (qCBV in ml/100g) using parenchymal T1 changes resulting from the Gd administration using a two compartment model and included the effects of intra- to extra-vascular water exchange[17]. Deconvolving the voxel-wise parenchymal tissue curves against the Global-AIF yielded mean transit time (MTT) values for the calculation of qCBF using the central volume principal,

$$qCBF(ml/100g/min) = \frac{qCBV(ml/100g)}{MTT(min)}.$$

In the remainder of this manuscript, we refer to these values as "Global-AIF perfusion" with Tmax calculated as the time at which the residue function reaches its maximum after deconvolution with the Global-AIF[24].

**Local-AIF DSC Post-Processing:** In a setting of vascular occlusion, the traditional Global

AIF chosen as a global estimate of the contrast bolus shape as it enters the brain has been shown to overestimate degree of hypoperfusion in vascular beds fed via collateral blood supply[14]. To address this shortcoming, a Local-AIF was automatically generated for every voxel within the deconvolution analysis of DSC perfusion to yield a map of perfusion in ml/100g/min. This Local-AIF was calculated using a previously reported technique [10],

$$AIF_{local} = AIF_{global}(t - \Delta t) \otimes \frac{\alpha}{(\Delta t + 1)} e^{-\frac{\beta t}{\Delta t}}$$

which includes delayed arrival time ($\Delta t$) and bolus dispersion ($\otimes e^{-\frac{\beta t}{\Delta t}}$) with β determined individually from comparison to between Global AIF and venous outflow. This algorithm for corrects for delay and dispersion by automatically generating a local input function for each voxel and has been validated through direct comparison with $H_2[O^{15}]$ PET in humans[10]. The Local-AIF is patient dependent and accounts not only for the delay of the arrival of the contrast bolus and its flow through the collateral network, but also the dispersion of the bolus shape (i.e., broadening) prior to its arrival at the vascular bed in question. The Local-AIF accounts for both antegrade flow through a tight flow-limiting stenosis, cross-filling through Circle of Willis communication, and retrograde blood supply through collateral vascular networks when calculating perfusion.

**Territories of Interest:** Three physiologically relevant territories were operationally defined: (1) diffusion positive core (core infarct), (2) hypoperfused yet viable (penumbral tissue-at-risk) and (3) contralateral normal hemisphere. All territories were defined based on mean diffusivity from DTI and Tmax calculated from DSC images using a Global AIF.

Core infarct was defined, as previously reported [11,12] as mean diffusivity below $5.7e^{-4}$ mm$^2$/s[6], converted to binary infarction maps, and used to calculate infarct volumes. Infarction maps taken 2.5 hours post-MCAO were co-registered to DSC (MATLAB 2021b, The MathWorks, Natick, MA) for analysis of the perfusion deficit in and around the core infarct.

Tissue at risk (i.e. the ischemic penumbra) was defined as having prolonged Tmax relative to the Global-AIF (Tmax > 1.0s), but not yet infarcted (i.e., MD>$5.7e^{-4}$ mm$^2$/s). Tmax[25,26] thresholds for "hypoperfusion" established in larger, human brains may not be appropriate in canines. Therefore, this study's generous threshold of Tmax>1.0s encompassed all tissue potentially at risk with delayed arrival. These Global-AIF Tmax maps, minus the binary infarction maps, were applied to the delay and dispersion corrected images as binary mask "tissue at risk" maps. Contralateral middle cerebral artery territory was selected by anatomic ROI. Perfusion beds in the ipsilateral hemisphere that were not "at risk" or core-infarct were not evaluated in analysis.

**Statistical Analyses:** Wilcoxon signed-rank, boxplots, and scatterplots (linear regression and Bland-Altman) were used to compare tissue perfusion as calculated with a Global-AIF and Local-AIF. Local-AIF DSC, Global-AIF DSC, and the difference between the two (ΔqCBF= $CBF_{LOCAL-AIF}$ - $CBF_{GLOBAL-AIF}$) were correlated against reference standard x-ray angiographic PCS with linear regression. To examine how relative CBF would be impacted by AIF selection, rCBF was calculated as the ratio of tissue at risk to contralateral middle cerebral artery territory. As robust collateral blood supply would maintain homeostasis and slow infarct growth, Local AIF perfusion and Global-AIF perfusion distal to the occlusion were correlated against infarct growth via linear regression. Local-AIF and Global-AIF qCBF were calculated in ml/100g/min within territories of interest (infarcted, tissue at risk, contralateral) and compared via Wilcoxon signed-rank. Wilcoxon rank sum was used to compare Local-AIF and Global-AIF perfusion in tissue-at-risk for cases with poor collaterals against cases with good collaterals. To investigate potential influence of AIF selection and assess whether differences in CBF resulted from blood volume calculation versus transit time calculation, the difference in quantitative CBV (ΔqCBV=$qCBV_{Local-AIF}$-$qCBV_{Global-AIF}$) was compared as a function of collateral score.

The bootstrap z-test was used to estimate effect size for power calculation. All statistical analysis was performed in R (Rstudio 3.6.1, Posit PBC, Boston, MA USA, 2019) and Python 3.11.4 (Anaconda Inc., 2024) with statistical significance determined at the p = 0.05 level.

**RESULTS**

Access to all results, raw images and tabulated data is available upon request to the corresponding author. The overall success rate of the experiments under the parent study was 83% (this study analyzed only successful controls) which reached 100% success for the last nine consecutive experiments. Representative parametric perfusion images of a left MCAO calculated with a Global-AIF and a Local-AIF for subjects with (A, B) poor collateral supply (PCS=8) and (C, D) good collateral supply (PCS=11) are shown in Fig. 1. The observed differences in perfusion (A vs B, C vs D) result exclusively in post-processing from the choice of AIF (Global vs Local); identical DSC images were used for A-B and for C-D. The corresponding 1-hour infarct and 4-hour infarct of the two cases in Fig. 1 are shown in Fig. 2 as mean diffusivity images. Visual inspection of Fig. 1 demonstrates that the presence of robust collateral supply (PCS=11) effectively mitigates the formation and growth of an infarct (Fig. 2, yellow arrows).

**Collateral Supply:** Local-AIF perfusion values in the tissue-at-risk were more strongly correlated (Fig. 3A, $R^2 = 0.81$) to pial collateral score than traditional Global AIF DSC (Fig. 3B, $R^2 = 0.05$). Correlation of *relative* perfusion (Tissue-at-Risk /Contralateral) to PCS showed Local-AIF relative perfusion values ($R^2=0.62$) were again more strongly correlated to pial collateral score than Global-AIF ($R^2=0.04$). The difference between Local-AIF and Global-AIF perfusion, ($\Delta qCBF = CBF_{LOCAL-AIF}$ and $CBF_{GLOBAL-AIF}$) was moderately positively correlated with PCS ($R^2 = 0.49$).

**Infarct Growth:** Perfusion calculated using the Local-AIF algorithm in the tissue at risk was strongly negatively correlated with infarct growth (higher local perfusion = slower infarct growth) (Fig. 4A, $R^2$=.79). In comparison, perfusion values calculated using a single Global-AIF were consistently lower than Local-AIF values and did not correlate with infarct growth for the same groups (Fig. 4B, $R^2$=0.02).

**Comparison in Territories of interest:** A comparison between traditional Global-AIF and delay and dispersion corrected Local-AIF perfusion in ml/100g/min is shown as boxplots in Fig. 5. Median perfusion of the core infarct (Fig. 5A, $MD < 5.7 x 10^{-4}$ mm$^2$/s and Tmax > 1.0 s), tissue-at-risk (Fig 5B, $MD > 5.7 x 10^{-4}$ mm$^2$/s and Tmax > 1.0 s), and the contralateral hemisphere (Fig. 5C) are included. In the infarcted core both Global and Local AIFs showed tissue perfusion values below the historical threshold for cell death (gray band < 18 ml/100g/min)[27]. In the tissue-at-risk, the perfusion of "good" collaterals and "poor" collaterals is presented separately to highlight the sensitivity of the Local-AIF to collateral supply.

Statistical comparison with Wilcoxon signed rank are shown in Table 1. Median perfusion in ml/100g/min from Local-AIF and Global-AIF were not statistically significantly different in the core infarct. In the tissue at risk, Local-AIF and Global-AIF were not significantly different for poor collaterals. However, for good collaterals, Local-AIF was significantly higher than Global-AIF CBF by +26.0ml/100g/min (p = 0.06), raising the average CBF above the ischemic threshold. Further, the Local-AIF perfusion was an average +25.9ml/100g/min higher for subjects with good collaterals compared to Local-AIF of poor collaterals (rank sum statistic = -2.2, p=0.03) while Global-AIF CBF of good and poor collaterals was not significantly different.

In the contralateral hemisphere the Local-AIF CBF was lower for good collaterals by -92.5ml/100g/min (rank sum statistic = 1.9, p=0.05). In other words, Local-AIF showed both higher perfusion in tissue-at-risk and lower contralateral perfusion for subjects with robust collateral blood supply while Global-AIF showed lower perfusion in the tissue-at-risk and higher contralateral perfusion for good collaterals. As such, *relative* perfusion rCBF=tissue-at-risk/contralateral perfusion with a Local-AIF correlated to collateral score ($R^2=0.62$) while Global-AIF rCBF did not ($R^2=0.04$). The influence of Local AIF on qCBV ($\Delta qCBV[ml/100g]$ = $CBV_{LOCAL-AIF}$ and $CBV_{GLOBAL-AIF}$) was not significant.

**Statistical Power:** Regarding power analysis for sample size, difference in the slopes for prediction of infarct growth using Local-AIF (Fig. 4A) or Global-AIF (Fig. 4B) perfusion did not achieve statistical significance by the bootstrap z-test ($z=-0.49$, $p = 0.62$). However, $R^2$ values did reach statistical significance ($z=2.03$, $p = 0.04$). In terms of power, the study was underpowered for both endpoints due to the sample size, with a minimum requirement of 41 subjects for ideal 80% power of both slope and $R^2$. However, the trend of improved slope and correlation with a Local-AIF remains (Fig. 4).

## DISCUSSION

This study found that tissue perfusion calculated using a previously reported voxel-wise "Local-AIF" was a better predictor of tissue status than a traditional single Global-AIF. The Local-AIF perfusion values were found to reflect pial collateral score and infarct growth in ischemic stroke. These findings suggest that a traditional Global-AIF does not include compensatory collateral blood supply and may falsely report a greater volume of hypoperfusion when robust leptomeningeal collateralization exists. An automatic Local-AIF can incorporate delayed and dispersed bolus to reflect the compensatory blood flow in tissue distal to an occlusion. These findings support use of a Local-AIF to (1) calculate quantitative perfusion using the previously reported "T1 bookend method"[9,15-19], (2) image local perfusion supplied through the leptomeningeal arteries, and (3) more accurately predict of infarct growth.

Quantification of local CBF distal to an occlusion using a Local-AIF demonstrated a strong correlation with pial collateral score. Further, for subjects with good collaterals, Local-AIF CBF in the tissue at risk, which would benefit most from robust collateral supply, was raised above the ischemic threshold. Meanwhile, in the core infarct, and in tissue-at-risk for subjects with poor collaterals and minimal collateral supply, Local-AIF showed no change in the degree of hypoperfusion: the use of a Local-AIF was unnecessary.

Global-AIF perfusion demonstrated poor correlation to collateral supply, and no difference between good and poor collaterals, suggesting Global-AIF does not reflect the presence of delayed blood supplied through the leptomeningeal collateral network. The absolute difference between the Global-AIF perfusion and the Local-AIF perfusion (ΔqCBF) correlated with collateral score Local-AIF including degree of collateral blood supply (vs slow antegrade stenotic flow) in an ischemic stroke. As both Global- and Local- AIF use identical input images and regions of interest, the difference was only due to the selection of AIF. The effect of the Local-AIF was observed in relative perfusion as well. Local-AIF is an important correction in DSC of acute stroke to include compensatory collateral perfusion distal to an occlusion, even if the intent is for relative perfusion.

Prior studies have reported correlation between infarct growth from both pial collateral score and arrival time by DSC[13]. The current study strengthens prior studies by indicating that subjects with higher Local-AIF perfusion, which includes contribution from collateral supply and corrected arrival time, observed slower infarct growth. Local perfusion may be a predicter of infarct growth and provide more insight into method of action of pre-clinical stroke treatment[7].

Tmax, by definition, identifies the presence of delayed bolus arrival as prolonged Tmax and depicts malignant pattern used in acute stroke triage that identifies the need for immediate intervention[1]. The goal of this study was to quantify the leptomeningeal supply that mitigates infarct growth. The results suggests that Tmax can identify the location of delayed bolus arrival as tissue at risk, and that a Local-AIF can calculate the quantitative local flow (i.e. degree of supply) from the collateral network in that tissue at risk.

Previous study has reported "local perfusion fraction" in acute stroke using non-contrast IVIM, with excitation and readout in the same plane, as a measure of local collateral supply that traditional DSC cannot capture[28]. Our study demonstrated that traditional DSC perfusion can be localized by correcting for the delay and dispersion. The correlation of Local-AIF DSC perfusion to x-ray DSA pial collateral score supported the ability of DSC to quantify local perfusion as collateral supply in ml/100g/min. In addition, as the influence of a Local-AIF on qCBV was not significant, the increased correlation with a Local-AIF was predominately due to the mean transit time component rather than the blood volume component. Since quantitative perfusion in ml/100g/min from a combination of the qCBV (with a T1 Bookend) and local mean transit time (with a Local-AIF) correlated with pial collaterals and infarct growth, studies of local IVIM perfusion fraction[28] may benefit from an inclusion of local mean transit time to capture collateral supply[29] as a method of calculating collateral supply.

This work was not without limitations. Use of a complex pre-clinical animal model limited the number of experiments that were run and vascular territories that were studied. We present a retrospective re-analysis of an animal model that designed to study flow augmentation in MCAO and was not powered for study of more subtle Local- vs Global-AIF perfusion differences between groups. This MCAO model and use of contrast afforded a limited number of perfusion measurements to assess dynamic changes in perfusion over the course of infarct development. A method that allows dynamic local perfusion could be beneficial. Co-registration and resampling were required for mapping of core-infarct and tissue-at-risk due to DSC and DTI having different FOV and dimensions which resulted in imperfect co-registration of perfusion and diffusion images. We therefore analyzed our images region-by-region, rather than voxel-by-voxel. Translation of conclusions derived from animal-based models is a potential limitation; however, an MCAO model reduces error in evaluating methods for CBF calculation by allowing from more accurate assessment of occlusion time and comparison of subjects at multiple time points under similar physiologic conditions which is not possible in humans. The limited number of subjects used in this study may have made it less likely to detect a statistically significant differences, especially in the expected smaller volume and higher relative variability of tissue at risk in subjects with poor collaterals relative to larger volumes of tissue at risk in subjects with good collaterals.

**CONCLUSION**

By using Local-AIF DSC to calculate tissue perfusion, the blood flow supplied by pial collateral recruitment was quantified, correlated with pial collateral score, and predictive of infarct growth. These findings support use of a Local-AIF rather than a single Global-AIF to (1) calculate perfusion in tissue-at-risk, (2) reduce false hypoperfusion in DSC of large vessel occlusions, and (3) quantify the blood supplied to a compromised perfusion bed through the leptomeningeal arteries. This quantitative Local-AIF DSC perfusion may be of benefit to studies of novel stroke therapeutics that would be influenced by leptomeningeal collateral supply.

**Data Availability:** All data is available upon request to the corresponding author.

Table 1: Differences between Local-AIF and Global-AIF DSC (ΔqCBF) in the three territories

**ΔqCBF DICHOTOMIZED BY COLLATERAL SCORE**

**MEAN DIFFERENCE (ML/100G/MIN) [WILCOXON SIGNED RANK STATISTIC, P-VALUE]**

|  | Core Infarct | Tissue at Risk | Contralateral Hemisphere |
|---|---|---|---|
| **ALL PCS** | +2.26[11.0, p = 0.68] | +14.3[5.0, p = 0.07] * | -30.06[16.0, p = 0.84] |
| **GOOD PCS>8** | +10.78[0.0, p = 0.125] | +25.96[0.0, p = 0.06] * | -89.2[2.0, p = 0.19] |
| **POOR PCS<8** | -9.08[1.0.0, p = 0.50] | -5.07[1.0, p = 0.50] | +68.60[1.0, p = 0.50] |

of interest, with all collateral scores dichotomized into good and poor. Wilcoxon signed rank statistic comparing the local-AIF DSC and the Global-AIF DSC that make up the ΔqCBF is included. Statistical significance at the p = 0.10 level is denoted with *. The underestimation of perfusion values using a single Global AIF compared to a voxel-wise local AIF (ΔqCBF) is greater when collateral supply is robust but not significant in a setting of poor collateral supply; this is especially prevalent in tissue-at-risk. DSC: dynamic susceptibility contrast, qCBF: quantitative cerebral blood flow, PCS: pial collateral score

**FIGURES AND TABLES**

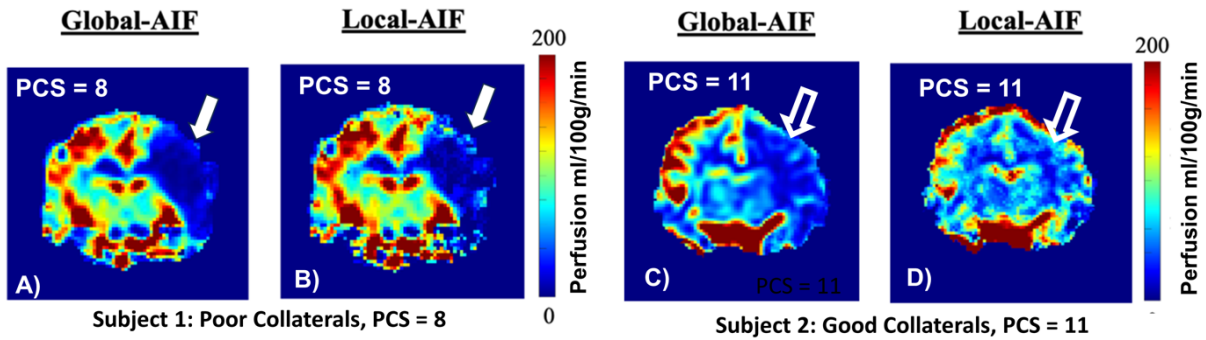

Figure 1: Coronal projection parametric perfusion images in ml/100g/min. A) Poorly collateralized subject (PCS=8), using a single Global AIF to calculate perfusion. B) The same subject and identical DSC images using voxel-by-voxel Local-AIF to calculate qCBF. Note the severe hypoperfusion (deep blue, denoted by arrow) in the poor collateral supply case, which persists when using a Local AIF. C) A subject with robust collateralization (PCS =11) shown with Global-AIF used to calculate perfusion and D) the same subject/images using Local-AIF to calculate perfusion. Note the markedly higher perfusion values (hollow, white arrow) attributable to the use of a delay and dispersed, Local AIF.

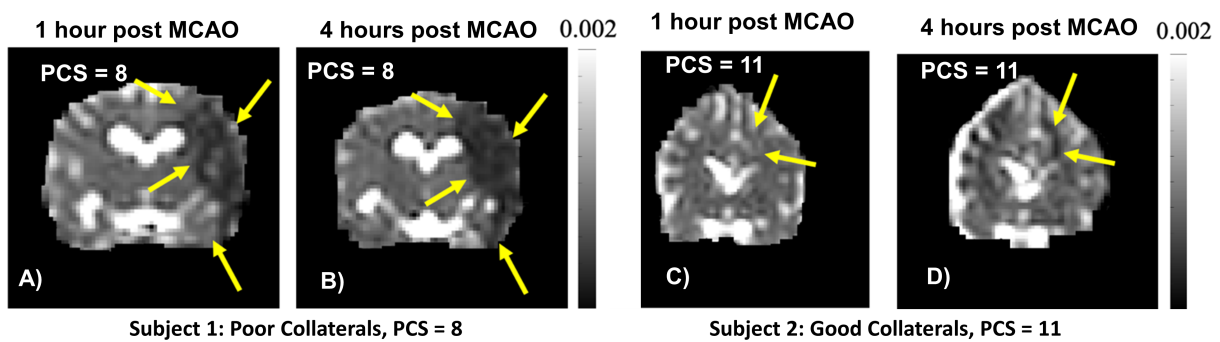

Figure 2: Coronal projection parametric mean diffusivity images acquired 1 and 4 hours after coil deployment in the left MCA. These are corresponding diffusion weighted images from the experiments displayed in Fig. 1 with A-B) poor collateral supply and C-D) good collateral

ARXIV PREPRINTsupply. Note that robust collateralization results in significantly slower growing, and smaller infarction (yellow arrows).

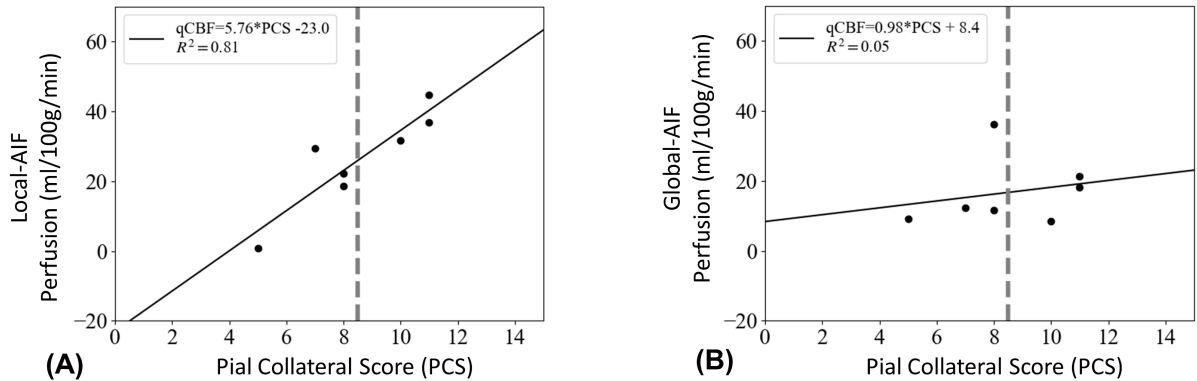

Figure 3: Scatter plots of tissue perfusion in ml/100g/min vs pial collateral score (PCS) within the operationally defined ischemic penumbra reconstructed with (A) a Local-AIF and (B) a traditional Global-AIF. Note: perfusion values are quantitative, values above ~18 ml/100g-min have sufficient flow to slow infarct growth.

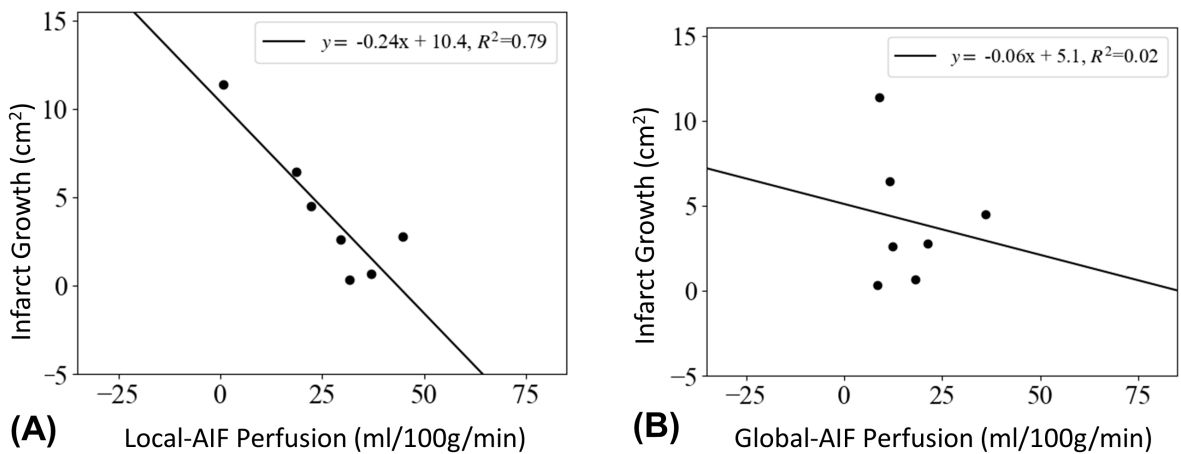

Figure 4: In the ischemic penumbra (MD<$5.7x10^{-4}$ mm$^2$/s, Tmax>1.0) A) Tissue perfusion as calculated using a Local-AIF more closely correlates with slower infarct growth (black markers, black line) whereas B) a traditionally chose Global-AIF images are less predictive of infarct growth in the acute phase of an ischemic stroke.

Arxiv.org PREPRINT 06/06/2024                                                                                                   19

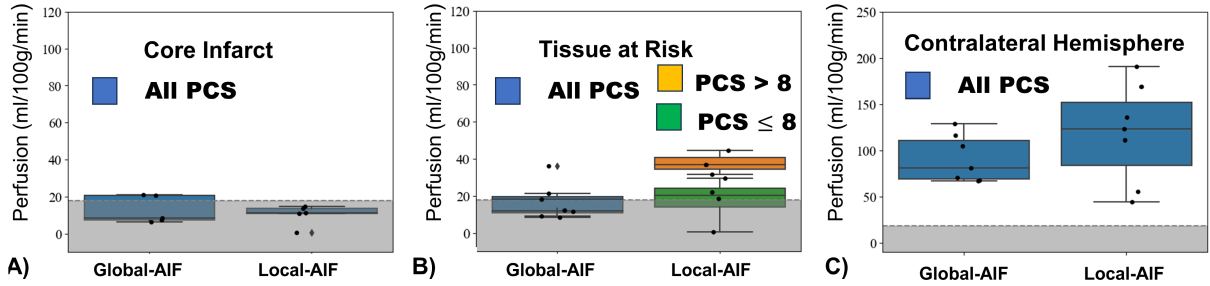

Figure 5: Boxplots of perfusion in A) the core of an infarct, B) the penumbral tissue-at-risk, and C) contralateral, unaffected hemisphere. Local-AIF DSC perfusion in tissue-at-risk is separated into good collaterals (PCS>8) and poor collaterals (PCS<8). Contralateral perfusion is on a different range due to systemic compensatory mechanisms to include all data. The gray band represents perfusion values below 18 ml/100g/min, the ischemic threshold for neuronal death.